\documentstyle[aps,prc,twocolumn,psfig]{revtex}
 
\newcommand{\eos}{equation of state~}

\newcommand{\eosp}{equation of state}

\newcommand{\beqn}{\begin{eqnarray}}
\newcommand{\eeqn}{\end{eqnarray}}

\begin{document}
\tightenlines
\title{Non-Identical Neutron Star Twins}
\author{Norman K. Glendenning}
\address{
Nuclear Science Division, Lawrence Berkeley National Laboratory,\\
University of California, Berkeley, CA 94720, USA}
\date{June 17, 1998}
\author{Christiane Kettner}
\address{Institut fuer theoretische Physik I,
Universitaet Augsburg\\
Memmingerstr. 6,
86135 Augsburg}
\maketitle

\newcommand{\doe}
{This work was supported by the
Director, Office of Energy Research,
Office of High Energy
and Nuclear Physics,
Division of Nuclear Physics,
of the U.S. Department of Energy under Contract
DE-AC03-76SF00098.}

\begin{abstract}
The work of J. A. Wheeler in the mid 1960's 
showed that for  smooth equations of state
no stable stellar configurations  with central densities
above that corresponding to the limiting mass of ``neutron
stars'' (in the generic sense) were stable against acoustical vibrational
modes. A perturbation would cause any such star to  collapse to a black hole
or explode. Accordingly, there has been no
reason to expect that a  stable degenerate family of stars with higher 
density than the known white dwarfs and neutron stars might
exist.
We have found a class of exceptions corresponding to certain  equations
of state 
that describe  a first order phase transition.
We discuss how such a
higher density family of stars could be formed in nature,
and how the promising new exploration of oscillations in the X-ray
brightness of accreting neutron stars might provide a means of identifying
them. Our proof of the possible 
existence of a third family of degenerate stars is one of principle and rests
on
general principles like causality, microstability
of matter
and General Relativity.

\end{abstract}

~\\[.1ex]


Since the early work by Wheeler and collaborators
\cite{wheeler65:a}, there have been
two reasons to doubt that there are any families of stable degenerate
stars above the density of the known white dwarf and neutron star
families.
The  first reason
is a
physical one and can be understood as follows.
Concerning the class of degenerate stars,
white dwarfs are stabilized by degenerate
electron pressure which fails at such density that electron capture
reduces their effectiveness. Stability is reestablished at densities
about five 
orders of magnitude higher when the baryon Fermi pressure (and
ultimately the short-range nuclear repulsive interaction) supports neutron
stars.

There is no evident mechanism for stabilizing a denser family. At higher
density the Fermi pressure of nucleons and hyperons is replaced---not
supplemented---by the pressure of their quark constituents if the
principle of asymptotic freedom
 \cite{asymptotic} is correct. Indeed a phase transition 
will generally reduce the pressure at a given energy density, and 
ultimately bring 
about
the termination of stability to gravitational collapse. Moreover,
at ever higher
density the Fermi pressure of quarks must be shared among a greater number
of flavors according to the Pauli principle. 
A mechanism for stabilizing a third family is therefore
not apparent from the point of view of Fermi pressure.

\begin{figure}[b]
\vspace{-.4in}
\begin{center}
\psfig{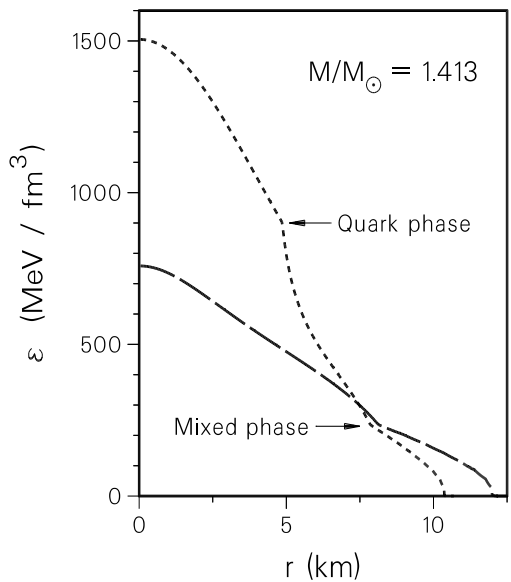}
\vspace{.3in}
\parbox[t]{4.6 in} { \caption { \label{prof_k290m66b180} Mass-Energy
profiles of ``twin'' stars of the same mass. Density thresholds are marked.
The high density
twin has a pure quark matter core extending to 5 km surrounded by
a mixed phase shell and that by a confined  hadronic phase shell. The normal
star has a mixed phase core extending to 8 km.
The pure quark phase is absent in the low-density twin.
}}
\end{center}
\end{figure}

The second reason for the  general belief that no stable stellar 
configurations exist with central densities higher than  those of 
neutron stars is due
to an analysis performed by 
Wheeler et al. Those authors showed analytically for  polytropic equations
of state that for general relativistic
stars there is a denumerable infinity of turning points
of the mass as a function of central density and therefore an infinity
of sequences for which the mass has positive slope.
 However, all configurations with densities greater than
the first mass limit for neutron stars
are unstable to acoustical radial vibrations, and end
either by exploding or imploding to a black hole. It seemed plausible that
this result was not peculiar to the polytropes, but would hold for any
(at least reasonably) smooth equation of state and there are 
numerical examples in the literature \cite{glen94:b} that demonstrate
that the theorem holds more generally. Moreover, and this is
important, the demonstration of instability of polytropes above the
neutron star family proves: (1) that 
positive slope of stellar mass with respect to central density
$dM/d\rho_c>0$---is  
not a sufficient condition for stability. (2) Therefore, the burden of
proof that any proposed third family of degenerate stars is stable
requires a demonstration that the normal modes of radial vibration of
stars in the family are stable rather than leading to collapse or explosion.

We have found exceptions to the expectation, corresponding 
 to  equations of state
that are physical in the sense that they are causal and microscopically
stable (Le Chatelier's principle)
but insufficiently smooth to obey the
quoted theorem. Under
certain combinations of parameters defining the nuclear and quark
deconfined equations of state, there exists a  sequence of high density
stellar configurations above the
neutron stars. The  ``neutron star'' sequence is
terminated by the softening in the equation of state by the mixed
phase when a substantial core of mixed phase is attained. A new sequence
at higher density is stabilized by replacement of the mixed phase
by  a pure quark phase core.
In this case there are stars
of the same
mass but radically different quark content and also of size. 

\begin{figure}[htb]
\vspace{-.4in}
\begin{center}
\psfig{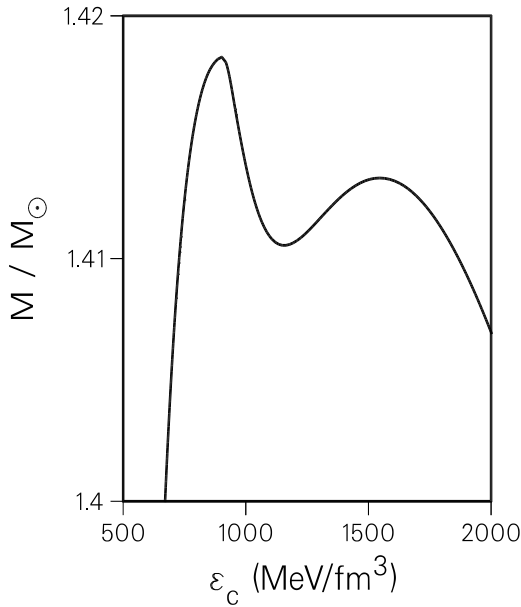}
\vspace{.3in}
\parbox[t]{4.6 in} { \caption { \label{MAS_K290M66B180} Stellar sequence
of neutron stars and their twins as a function of central density.
Both segments with positive slope correspond to stable
configurations since their normal modes of vibration are found to
be stable. 
}}
\end{center}
\end{figure}
We hasten to add however that what determines the 
stability of stars is the \eosp. Any other phase transition that has
similar effects on the \eos
as the class we have found could likewise give rise
to a third stable family. In any case we refer to the denser sequence
as non-identical twins of neutron stars because in both cases it is
the Fermi pressure of particles carrying baryon number that supports the
star against gravity in addition to repulsion at short distance
between any nucleons that may be present. We demonstrate that
in principle
it is permitted by the laws of nature that there can exist two
families of 
relativistic stars with partially overlapping regions of mass, for which
two `neutron' stars of the same mass exist, but in which baryon number
is distributed differently among the various species of baryons or
quarks.

An example of 
non-identical twins is shown  in Fig.\
\ref{prof_k290m66b180}
where we compare the density profiles.
The low-density star lies on the first segment of Fig.\ \ref{MAS_K290M66B180}
with positive slope and the high-density star lies on the second.
Details of the models of hadronic and quark matter can be found in
ref.\ \cite{book}. The nuclear matter properties that are used
to determine the hadronic equation of state
(beside the well established saturation density, binding and
symmetry energy coefficient) are the compression modulus
and effective nucleon mass at saturation, $K=290$ MeV and
 $m^\star/m=0.66$, respectively. The bag constant
for the quark equation of state is $B^{1/4}=180$ MeV and the quark model
equation of state is given in Ref. \cite{farhi84:a} with $\alpha_s=0$.

\begin{figure}[htb]
\vspace{-.4in}
\begin{center}
\psfig{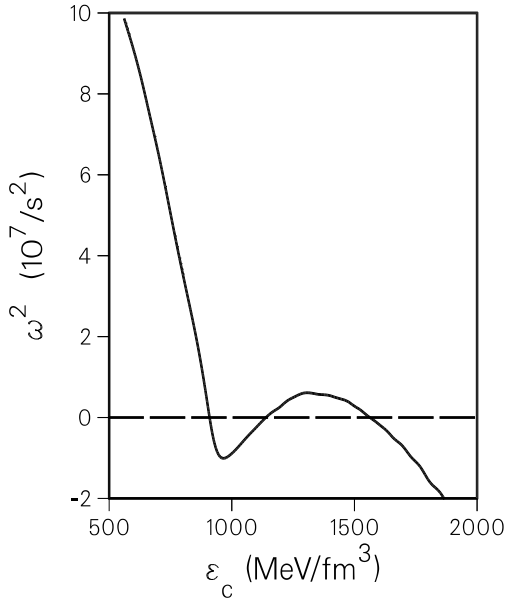}
\vspace{.3in}
\parbox[t]{4.6 in} { \caption { \label{kettner} The square of the frequency
of the fundamental radial vibration. Perturbations grow as $\exp(\omega t)$
when $\omega^2 < 0$. The segments with $\omega^2 > 0$ are stable
since the time dependence of a radial perturbation is $\exp(i \omega t)$.
}}
\end{center}
\end{figure}

Stars on both segments  of the stellar sequence shown in 
Fig.\ \ref{MAS_K290M66B180} having positive slope are stable.
However, positive slope $dM/d\rho_c>0$ is a 
necessary but not a sufficient condition for stability.
Usually, and in all cases tested heretofore, the second and all others
are unstable to
radial vibrations \cite{wheeler65:a,glen94:b}. 
Stability can be hinted at  from the behavior of the
mass-radius relation but a definitive  test
involves an analysis of the radial
modes of oscillation
\cite{chandrasekhar64:a}. The squared frequency $\omega^2$ of the
fundamental mode is plotted in Fig.\ \ref{kettner}. Positive values
indicate  stability
and correspond to the segments with positive slope
in Fig.\ \ref{MAS_K290M66B180}. The analysis shows that
the fundamental (nodeless $n=0$)
oscillation becomes unstable at the
first maximum, as usual, but unusually, stability of this mode
is regained at the
following minimum, to be lost again at the next maximum. The usual
pattern is that at the  maximum in the neutron star family the
fundamental mode becomes unstable, and at each higher
minimum and maximum a higher mode in order $n=1,2,3\cdots$ becomes unstable.

\begin{figure}[htb]
\vspace{-.4in}
\begin{center}
\psfig{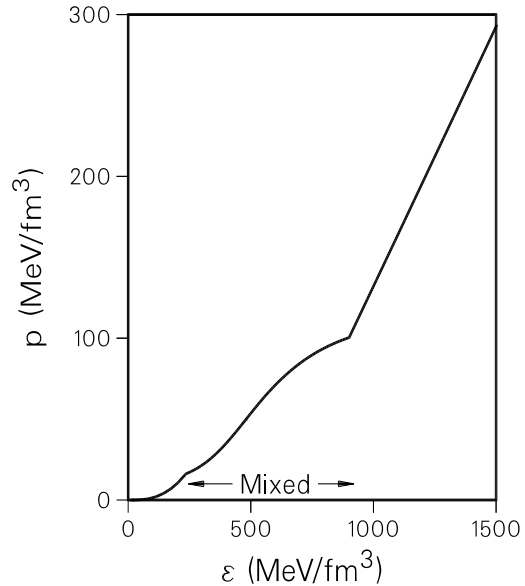}
\vspace{.3in}
\parbox[t]{4.6 in} { \caption { \label{EOS_K290M66B180} Equation
of state from the low
density normal nuclear matter through the mixed phase and into the
pure quark matter phase at high density
The phase transition is first order
and the pressure varies in the mixed phase because neutron star matter
has two independent components, baryon and electric charge
\protect\cite{glen91:a}. (The normal density of nuclear matter is
$140 {\rm~MeV/fm^3}$.)
}}
\end{center}
\end{figure}

We used the equilibrium equation of state
shown in Fig.\ \ref{EOS_K290M66B180} for the calculation of the oscillation
modes.
We do not know the time-scale for equilibration  of
the transition between quark matter and nuclear matter and how it compares
with   {\sl typical} periods of
pressure oscillations.
This does not matter. The turning points in the stellar sequence between
stability and instability occur only at the zeros of $\omega^2$, 
and because the stellar 
oscillations are infinitely slow at the turning points
their location  can be determined exactly.
Thus, independent of the time-scales, the stability for all
members of the stellar sequence can be determined.

The features of the stellar sequence  shown in Fig.\
\ref{MAS_K290M66B180}
can be identified with features in the equation of state
shown in Fig.\ \ref{EOS_K290M66B180}. The limiting mass star at $\approx
1.418 M_\odot$ has a central density ($\approx 900 {\rm~MeV/fm^3}$)
falling  very near the upper end of the
mixed phase where all quarks become deconfined.
 One can see from the equation of state near
 the end of the mixed phase that
 $dp/d\epsilon$ becomes small and therefore also  the
 adiabatic index,
 $\Gamma=d \ln p/d \ln \rho=(p+\epsilon)/p \cdot dp/d\epsilon$ (where
 $p,~\epsilon,{\rm~and~}\rho$ denote pressure, energy density and baryon
 density).
In this upper region of the mixed phase,
 the pressure exerted by matter is
insufficient to prevent collapse.  
The adiabatic index is discontinuous at the density
boundary between the mixed and pure  phases, being smaller in the 
mixed than in the pure phases. This is a characteristic of
phase equilibrium. The larger adiabatic index
in the pure quark phase restores
stability  over a short range of central densities
between
$\sim 1150 {\rm~MeV/fm^3}$ and  $\sim 1550  {\rm~MeV/fm^3}$,
above which the mass  becomes too large to be supported by the
Fermi pressure of quark matter and the second sequence of stable stars
terminates.

Twins,  found here for the deconfinement transition, are likely to
be a more general  phenomenon for  any first
order phase transition 
 so long as the low-density
 sequence of stars terminates at central densities that fall close to
 to the end of the mixed phase
 which in neutron star matter will have the
form shown in Fig.\ \ref{EOS_K290M66B180}.
(Generally,  the pressure
increases monotonically on an isotherm as in the figure
for a first order phase transition in a substance having more
than one conserved charge
\cite{glen91:a}.)

From the above proof of stability we have shown that in principle
a third family of stable degenerate stars could exist that lies in density
above the white dwarfs and neutron stars. 
In such a case, 
stars in the third family  could exist
whose masses match a small range of  normal neutron stars. 
Each star on the high density branch has a non-identical twin
on the low-density branch, having  the same
mass but different composition and radius.

How could a high density twin be made in nature? 
The likely path to the high density twins is through
 the initial core collapse of a star, in which the core implodes
through the normal star to the high density twin.  Since no two supernova
are likely to be identical, there being many variables that effect the
outcome, like mass, rotation, symmetry,  chemical composition of the
progenitor, and the chaotic process of convection, it seems plausible that
either twin could be produced.

A second possible formation mechanism of the high density twin is through
accretion onto a member of the low-density sequence of
Fig.\ \ref{MAS_K290M66B180}. This is perhaps the most interesting though 
not necessarily the most likely route.
Accretion onto a normal neutron star
near the mass  limit of the low-density
sequence could conceivably lead to collapse to a member of the high-density
sequence, with the excess mass being blown off in a minor nova
by the release
of gravitational binding energy, much
as it is thought that some white dwarfs may collapse by accretion to
a neutron star.

Can such twins be identified? One possible avenue is through
observations
on the so-called quasi-periodic oscillations in the X-ray brightness
of accreting neutron stars. According to theory, mass and radius
determinations may be possible \cite{lamb98:a,lamb98:b}.  If twins exist, then
the mass-radius curve will exhibit two segments of stable stars
instead of one,
and 
observed stars will fall on one or the other of the two distinct segments.
The discovery of only about two stars on each branch with a radius
resolution
of  a kilometer in our example would 
suffice to
establish the existence of twin branches.

Are there even higher density families of relativistic
compact stars? It is not ruled
out by virtue of the proof given here that a third family is allowed
by reason of structure introduced into the \eos by a phase transition.
There could be a series of phase transitions following each other
at a succession of densities. But the quark 
deconfinement phase transition may be the ultimate attainable
in compact stars that are stable to collapse to black holes.
At such densities that matter is in the pure quark phase, the
\eos is likely to be  smooth like a polytrope. From that
density on we are assured of  the denumerable infinity of
turning points in the stellar sequence
having an  increasing number of unstable normal
modes such as was found by Wheeler et.\ al.\ \cite{wheeler65:a}.

Our proof that a third family of degenerate stars is allowed by the laws of
nature is not a proof of their actual existence, quite aside from questions
concerning the
mode of formation.
From the discussion in the introduction, it is not a question of new
degrees of freedom being engaged. It is a question of the behavior of the
\eos of superdense matter. We have no knowledge from experiment of  a single
point on the
\eos above nuclear density. But we do have expectations of phase transitions,
and asymptotic freedom of quarks would appear to elevate one of them
to a law of nature. A phase transition can produce the requisite structure 
discussed above into the \eos so as to restore stability for a
finite density range after stability has been lost by the canonical
neutron stars. The possible existence of a third family of compact stars
hinges on such details that we may never determine in laboratory experiments.
The actual discovery of members of the third family therefore would
reveal, however imperfectly, a non-smooth behavior of the \eos
possibly caused by a change in phase of dense matter---that we may never
know by laboratory experimentation.

{\bf Acknowledgements:} \doe

\vspace{-.2in}


\end{document}